\documentclass[conference]{IEEEtran}
\usepackage{times}
\usepackage{fullpage}
\usepackage{graphicx}
\usepackage{subfigure}
\usepackage{flushend}
\usepackage{multirow}
\usepackage{multicol}
\usepackage[ruled]{algorithm2e}
\usepackage{algpseudocode}   
\usepackage{amsmath}
\usepackage{amsfonts}
\usepackage{amsthm}
\usepackage{pgfplots}
\DeclareMathOperator*{\argmin}{arg\,min}
\theoremstyle{rule}
\newtheorem{ruler}{Rule}[section]
\usepackage{url}

\SetAlFnt{\small}
\SetAlCapFnt{\small}
\SetAlCapNameFnt{\small}
\SetAlCapHSkip{0pt}

\begin{document}

\title{DeepObfuscation: Securing the Structure of Convolutional Neural Networks via Knowledge Distillation}

\author{\IEEEauthorblockN{
Hui Xu\IEEEauthorrefmark{1},
Yuxin Su\IEEEauthorrefmark{1},
Zirui Zhao\IEEEauthorrefmark{1},
Yangfan Zhou\IEEEauthorrefmark{2},
Michael R. Lyu\IEEEauthorrefmark{1},
Irwin King\IEEEauthorrefmark{1}}

\IEEEauthorblockA{\IEEEauthorrefmark{1} Dept. of Computer Science and Engineering, The Chinese University of Hong Kong}
\IEEEauthorblockA{\IEEEauthorrefmark{2} School of Computer Science, Fudan University}
}
\maketitle

\begin{abstract}
This paper investigates the piracy problem of deep learning models.  Designing and training a well-performing model is generally expensive.  However, when releasing them, attackers may reverse engineer the models and pirate their design.  This paper, therefore, proposes \textit{deep learning obfuscation}, aiming at obstructing attackers from pirating a deep learning model.  In particular, we focus on obfuscating convolutional neural networks (CNN), a widely employed type of deep learning architectures for image recognition.  Our approach obfuscates a CNN model eventually by simulating its feature extractor with a shallow and sequential convolutional block.  To this end, we employ a \textit{recursive simulation} method and a \textit{joint training} method to train the simulation network.  The joint training method leverages both the intermediate knowledge generated by a feature extractor and data labels to train a simulation network.  In this way, we can obtain an obfuscated model without accuracy loss.  We have verified the feasibility of our approach with three prevalent CNNs, \textit{i.e.,} GoogLeNet, ResNet, and DenseNet.  Although these networks are very deep with tens or hundreds of layers, we can simulate them in a shallow network including only five or seven convolutional layers.  The obfuscated models are even more efficient than the original models.  Our obfuscation approach is very effective to protect the critical structure of a deep learning model from being exposed to attackers.  Moreover, it can also thwart attackers from pirating the model with transfer learning or incremental learning techniques because the shallow simulation network bears poor learning ability.  To our best knowledge, this paper serves as a first attempt to obfuscate deep learning models, which may shed light on more future studies.
\end{abstract}

\maketitle
\thispagestyle{empty}

\section{Introduction}

We are experiencing a booming development of deep learning technologies.  Nowadays, more and more systems employ deep learning models, such as self-driving cars~\cite{bojarski2016end} and face recognition systems~\cite{parkhi2015deep}.  Karpathy even proposes the concept of \textit{Software 2.0}~\cite{software20} referring to the software written in neural network weights.  If we treat deep learning as a new paradigm of programming, it actually suffers many software security and reliability issues.  For example, DeepXplore~\cite{pei2017deepxplore} studies software testing approaches for deep learning models, because there are already several attacks (\textit{e.g.,}~\cite{nguyen2015deep,papernot2017practical}) showing the effectiveness in fooling them.  Another study~\cite{tramer2016stealing} shows that attackers can steal machine learning models via prediction APIs.  Therefore, the security of deep learning techniques becomes an urgent issue when being deployed in real-world systems.

In this paper, we consider a particular security threat to deep learning, \textit{model piracy}.  Programming a superior deep learning model is expensive.  It requires much domain expertise to design an effective deep learning network and a large set of labeled data to train the network, both of which are valuable resources.  Because well-trained models are expensive, competitors or attackers may get interested in pirating them.  An intuitive way is to copy the architecture of a network, which dominates the learning ability of a model.  Besides, attackers may fine-tune a model for their own application scenarios.  There are already many investigations focusing on building new deep learning applications based on existing ones, such as those with transfer learning~\cite{ganin2015unsupervised} and incremental learning~\cite{xiao2014error} techniques.  If a deep learning model runs on the client side, which is a trend (\textit{e.g.,} smartphone~\cite{howard2017mobilenets,lane2016deepx,wang2016accelerating}), attackers can easily reverse engineer the model and further pirate the design.  Therefore, model piracy is a pressing security concern for deep learning application providers.

To secure a deep learning model against piracy attack, we propose to obfuscate the structures of well-trained deep learning models before releasing them to clients.  Our idea is similar to classic code obfuscation techniques except that we tailor the idea for deep learning scenarios.  Code obfuscation transforms code snippets into unintelligible versions while preserving their semantics~\cite{collberg1997taxonomy}.  Deep learning obfuscation, on the other hand, aims to scramble the structure of a well-designed deep learning network while preserving the inference accuracy.  In this way, users can still employ an obfuscated model for inference, but attackers cannot learn useful structural information from the model.  To our best knowledge, this paper serves as a first attempt to study the deep learning obfuscation problem.

Our study focuses on a prevalent type of deep learning networks, convolutional neural networks (CNN).  Many companies start to engage CNN in their systems for image recognition tasks, \textit{e.g.,} in mobile apps or auto-driving systems.  To achieve a good recognition accuracy, state-of-the-art CNNs generally contain well-designed inception blocks for feature extraction and a fully-connected layer for classification.  For example, GoogLeNet~\cite{szegedy2015going} employes four parallel convolutional sequences in one inception block with different settings to learn different features; ResNet~\cite{he2016deep} employs a special convolutional branch to learning residual information.  For modern CNNs, their key difference generally lies in the design of the feature extractor, while the classifiers of different CNNs are very similar.  Therefore, hiding the real structure of a feature extraction network is a major concern for obfuscation.

We propose to obfuscate the feature extractor of a CNN model by simulating it with a shallow and sequential convolutional block.  Consequently, the simulation network leaks little structural information about the original feature extractor.  Meanwhile, the obfuscated model should also be resilient to fine-tuning attacks because the simulation network bears poor learning abilities due to a shallow structure.  To simulate a feature extraction network precisely, we incorporate a novel \textit{recursive simulation} method and a \textit{joint training} method to training the simulation network.  The recursive simulation method simulates a feature extractor in a recursive mode.  In the first round, we simulate each inception block of a feature extractor with a simulation network.  In the second round, we simulate the entire simulated feature extractor achieved in the first round.  During each iteration of simulation, we employ the joint training method to train a simulation network, \textit{i.e.,} we employ both the intermediate output of the original network and the labels of the training data as the ground truth.  Finally, we can obtain an obfuscated model with no loss of accuracy.  

To verify the feasibility of our idea, we have conducted several real-world experiments with popular CNNs, including GoogLeNet~\cite{szegedy2015going}, ResNet-18~\cite{he2016deep} and DenseNet-121~\cite{huang2017densely}.  Although their feature extraction networks are very deep with tens or even hundreds of layers, we can simulate them with a shallow network of five or seven layers.  We show that the obfuscated models suffer no loss of accuracy.  On the other hand, they are even more efficient than the original models in both model size and inference time.  We further show that the obfuscated models demonstrate promising resilience to fine-tuning attacks.  Attackers would suffer obvious accuracy declination if they fine-tune the obfuscated models to create new applications.

To summarize, we make several contributions as follows.

\begin{itemize}
\item We formulate the deep learning obfuscation problem with respect to model piracy.  In particular, we observe the potential \textit{structure piracy} and \textit{parameter piracy} threats to deep learning models and propose five metrics to evaluate a deep learning obfuscation solution, namely \textit{cost}, \textit{information leakage}, \textit{fine-tuning ability}, \textit{resilience} to deobfuscation attacks, and \textit{scalability}.

\item We propose a novel solution to obfuscate CNN models with \textit{recursive simulation} and \textit{joint training}.  Our approach can simulate the feature extractor of a CNN model with a shallow convolutional block, which conceals the structural information of the original network and also deters attackers from fine-tuning an obfuscated model.

\item We have verified the feasibility of our approach with several real-world experiments.  Our resulting obfuscated models suffer no loss of accuracy, and they are even more efficient than the original models in both model size and inference time.
\end{itemize}

We organize the rest of the paper as follows.  Section~\ref{sec:preliminary} briefly reviews the background of CNN.  Section~\ref{sec:attacker_model} defines our attack model for model piracy attack.  Section~\ref{sec:obfuscation} discusses the deep learning obfuscation problem.  Section~\ref{sec:approach} introduces our structural obfuscation approach, and the evaluation is provided in Section~\ref{sec:evaluation}.  Section~\ref{sec:related_work} compares our paper with related work, and Section~\ref{sec:conclusion} finally concludes this paper.

\section{Preliminary}~\label{sec:preliminary}
This section briefly reviews the techniques of CNN, which is a preliminary for the deep learning obfuscation approach we propose in this work.

\subsection{CNN Basis}
CNN is a special type of deep neural networks that contains convolutional layers and fully-connected layers.  The convolutional layers serve as a feature extractor of the network, which inputs images and outputs features.  The fully-connected layers serve as a classifier, which classifies images based on the extracted features. 

\begin{figure}[t]
\centering
\includegraphics[width=0.46\textwidth]{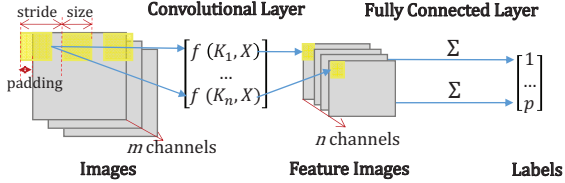}
\caption{A toy example of convolutional neural networks with one convolutional layer and one fully-connected layer.}
\label{fig:toy_cnn}
\end{figure}

\begin{figure*}[t]
\centering
\subfigure[An inception block of GoogLeNet.]{
\label{fig:googlenet_incep}
\includegraphics[width=0.34\textwidth]{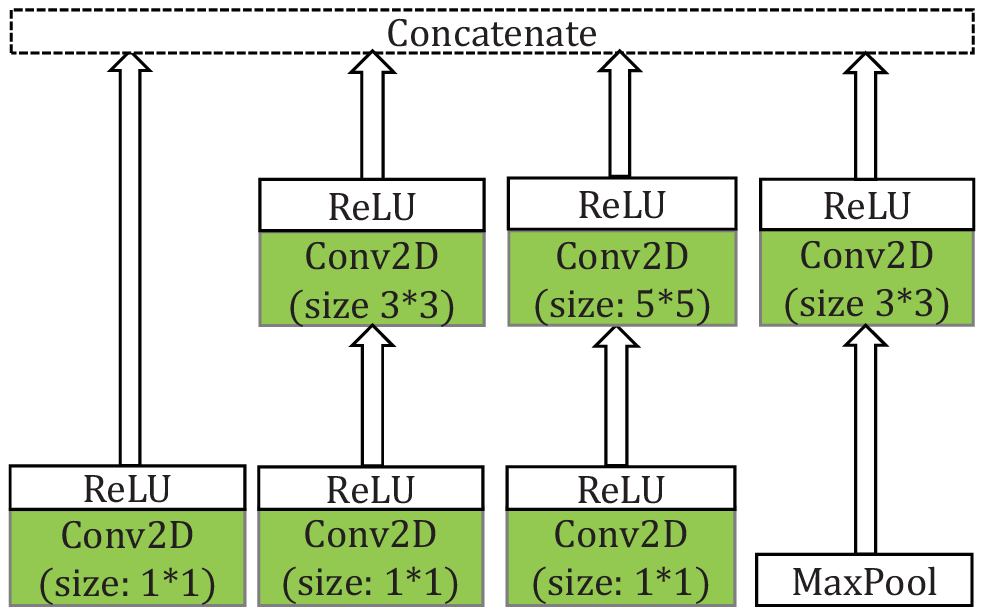}}
\subfigure[An sub-block of ResNet for inception.]{
\label{fig:resnet_incep}
\includegraphics[width=0.295\textwidth]{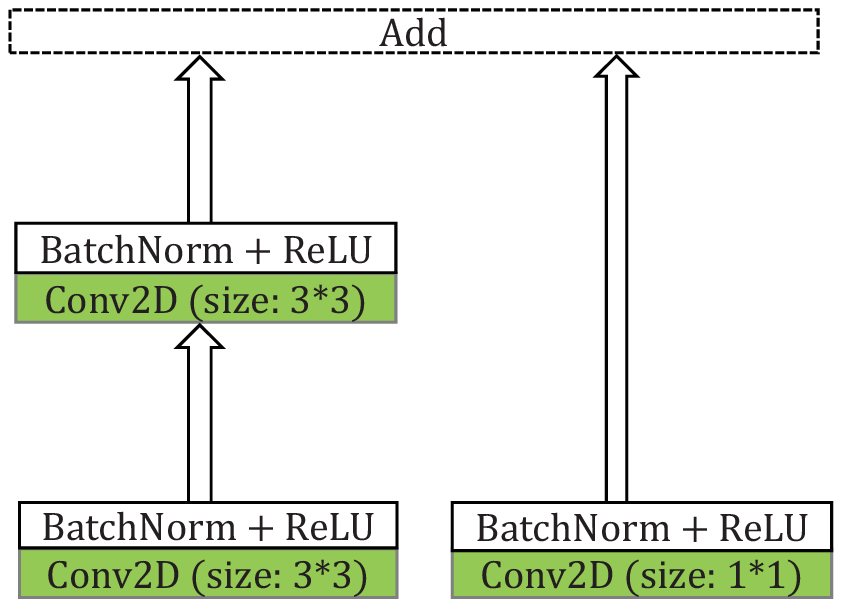}}
\subfigure[A partial dense block of DenseNet.]{
\label{fig:densenet_incep}
\includegraphics[width=0.3\textwidth]{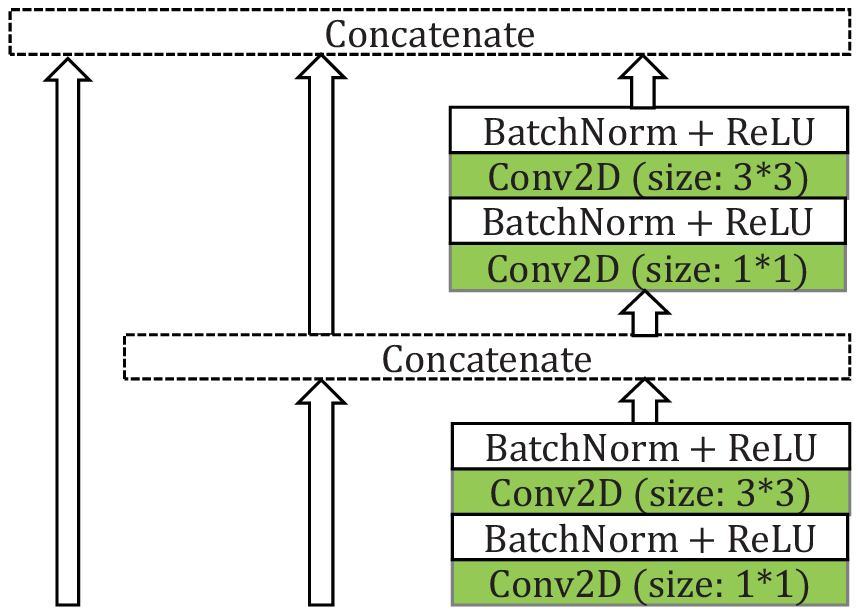}}
\caption{Examples of inception blocks.}
\label{fig:inception}
\end{figure*}

Figure~\ref{fig:toy_cnn} demonstrates a simple CNN with one convolutional layer and one fully-connected layer.  We discuss the detailed function of either layer in what follows.  We use uppercase letters to denote matrices and lowercase letters to denote the elements of a matrix.  For example, $K$ is a matrix, and $k_{i,j}$ is an element of the matrix.

The convolutional layer reads raw images $X\in \mathbb{R}^{h \times w \times m}$ with size $h \times w$ and $m$ channels (\textit{e.g.,} red, green, and blue for colorful images), and outputs $n$ feature images (\textit{i.e.,} $n$ channels). 
Each output channel corresponding to a convolutional kernel $K_i$ and a bias $\beta_i \quad \forall i\in\{1,\ldots ,n\}$ is calculated as follows. 

\begin{equation}
\mathcal{F}(K_i,X) = \sum_{j=1}^{m}{f(K_i,X_j)} + \beta_i,
\end{equation}
where $K_i$ is a matrix whose elements are the weights of corresponding pixels on an image $X$, and 
\begin{equation}
f(K_i,X_j) = \sum_{p=1}^{h}{\sum_{q=1}^{w}{k_{p,q} \cdot x_{p,q}}}.
\end{equation}

The fully-connected layer connects each pixel of each feature image to all the class labels.  A label with the highest value is voted as the final decision.  The formula for computing the value of each label is demonstrated in below.
\begin{equation}
G({X_1,...X_n}) = \sum_{i=1}^{n}{\sum_{j=1}^{h\times w}{x_{i,j}w_{i,j}+\beta_{i,j}}}.
\end{equation}

\subsection{Modern CNNs}
While conventional CNNs (\textit{e.g.,} LeNet~\cite{lecun1990handwritten}) only contain different layers organized in sequential orders, the architectures of modern CNNs are more complex.  They generally include well-designed inception blocks to facilitate the learning ability.  An inception block is a sub-network with convolutional layers and other nonlinear layers, such as batch normalization~\cite{ioffe2015batch} and ReLU (Rectified Linear Units).  These layers are similar to building blocks, and a programmer can organize them in many ways.

Figure~\ref{fig:inception} demonstrates the inception blocks of several popular CNNs, including GoogLeNet, ResNet, and DenseNet.  The inception block of GoogLeNet (Figure~\ref{fig:googlenet_incep}) contains four parallel sequences, each of which has a unique convolutional function to learn particular features.  All the convolutional sequences output feature tensors of the same size, and the inception block finally concatenates them as its output.  ResNet (Figure~\ref{fig:resnet_incep}) contains two parallel sequences in an inception block.  The sequence with a smaller convolutional kernel is designed to propagate the residual information.  The design is essential when a neural network goes deeper.  The inception block finally adds up the tensors outputted by the two convolutional sequences.  DenseNet (Figure~\ref{fig:densenet_incep}) further improves the mechanism for propagating residual information.  In each dense block, the output of every two convolutional layers is propagated to the following layers of the block.

A well-designed feature extractor is a key for a CNN to improve its performance when handling particular tasks, such as the ImageNet Large Scale Visual Recognition Challenge (ILSVRC)~\cite{deng2009imagenet}.  Note that all our discussed modern CNNs have only one fully-connected layer, and they mainly differ in the feature extraction network.  Therefore, hiding the structure of the feature extraction network is the most important concern when obfuscating deep learning models.

\section{Attack Model}~\label{sec:attacker_model}
\begin{figure}[t]
\centering
\includegraphics[width=0.46\textwidth]{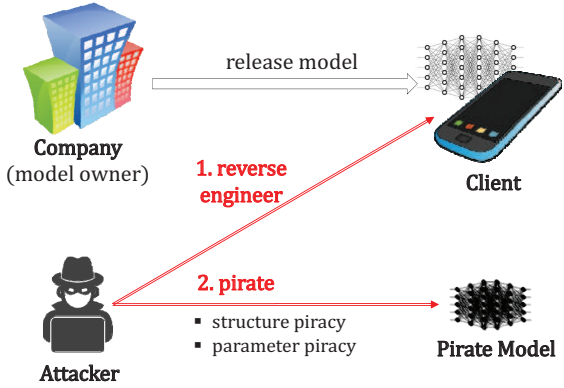}
\caption{Our assumed attack scenario.}
\label{fig:adversary}
\end{figure}

\begin{figure*}[t]
\centering
\subfigure[GoogLeNet (a tailored version with 1024 features) incremental learning with the CIFAR-100 dataset.  INC: incremental learning.]{
\label{fig:example_inc}
\includegraphics[width=0.47\textwidth]{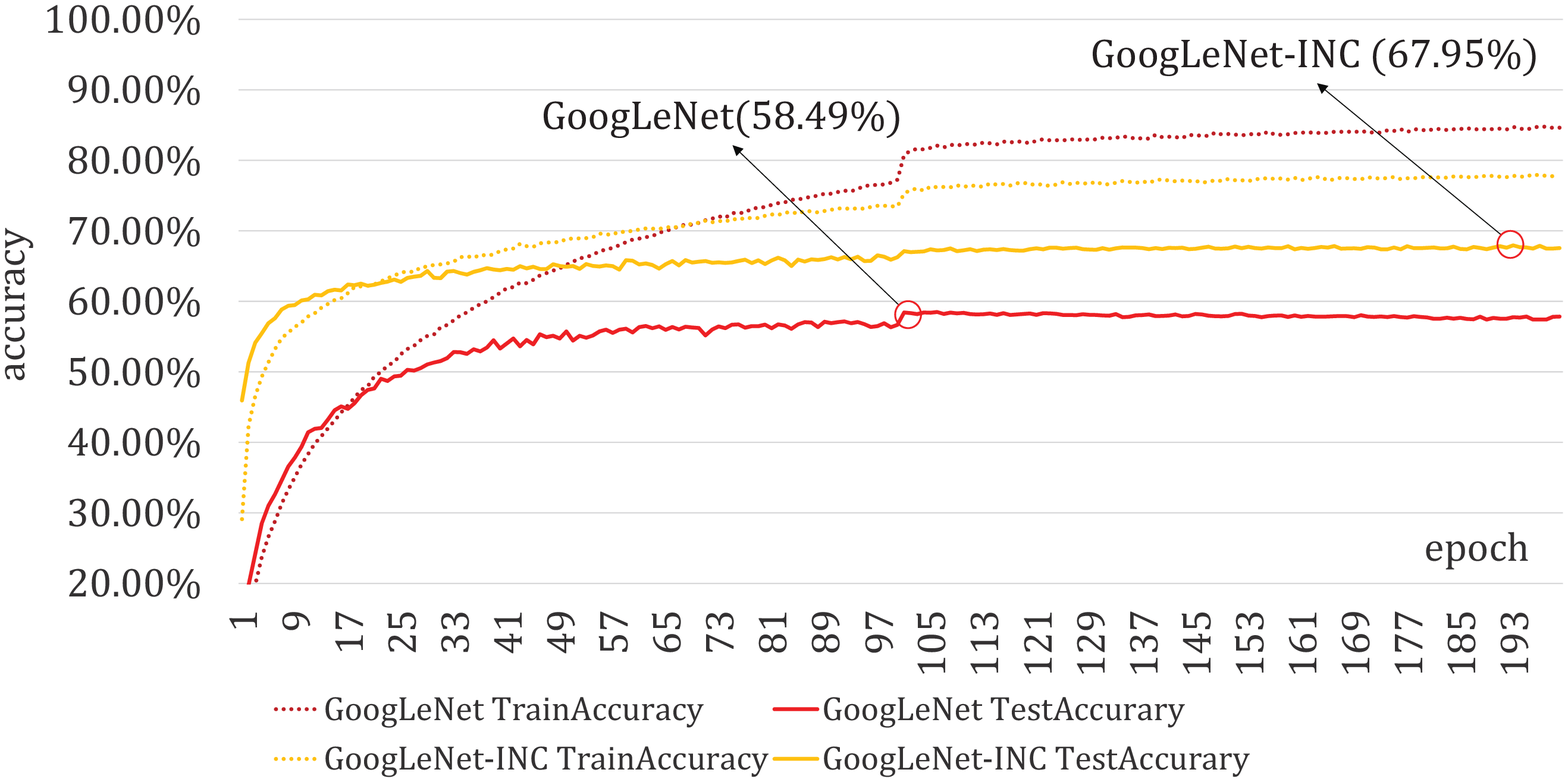}}
\hspace{0.5cm}
\subfigure[ResNet transfer learning with the STL10 dataset.  TLFF: transfer learning with frozen feature, TLFT: transfer learning with fine-tuning~\cite{pytorchtrans}.]{
\label{fig:example_trans}
\includegraphics[width=0.47\textwidth]{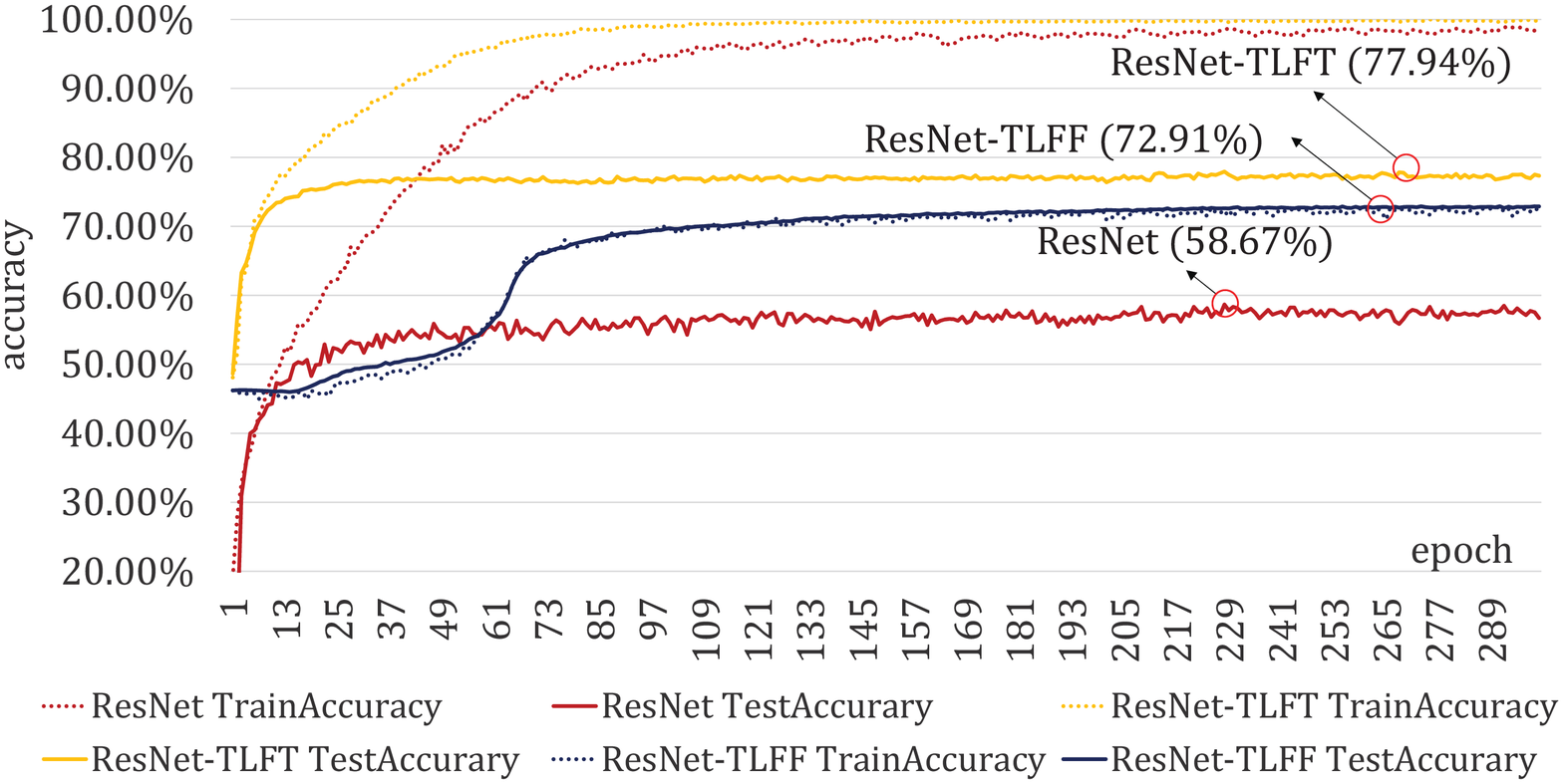}} 
\caption{Examples that demonstrate the value of parameter piracy attacks via fine-tuning. In our examples, the initial ResNet and GoogLeNet models are trained with the CIFAR-10 dataset.}
\label{fig:example}
\end{figure*}

This work considers the man-at-the-end (MATE) attack~\cite{collberg2011toward}.  Supposing a well-trained deep learning model has been installed on a client host (\textit{e.g.,} PC server or smartphone) for inference, the MATE attack model assumes that attackers can have full access to the host.  If attackers have adequate domain knowledge and determination, they may reverse engineer the application, and then they may pirate the design of the model.

Figure~\ref{fig:adversary} demonstrates our attack scenario.  We assume that a company (\textit{i.e.,} model owner) has released an application to its users, and the application contains a deep learning model.  We further assume that the model contains a novel design of network structure, and it is trained with a large corpus of valuable data.  The assumption is very general for real-world deep learning applications because such data are either the private assets of a company or require many  manual efforts to get labeled.  In the MATE attacker model, attackers (\textit{e.g.,} competitors) may reverse engineer the application and pirate the deep learning model.  Such infringement of copyrights would cause loss to the model owner.

In this work, we consider two types of piracy behaviors: \textit{structure piracy} and \textit{parameter piracy}.  In structure piracy, attackers reverse engineer an application and extract the model structure, and then they can employ the structure to design their own networks.  For CNN models, the distinctive design of structures lies in their feature extractors.  Therefore, an effective obfuscation solution should hide such structural information in the obfuscated model.  \textit{Parameter piracy} means attackers can employ the well-tuned model states (or parameters) to create new models, \textit{e.g.,} with \textit{incremental learning}~\cite{xiao2014error} and \textit{transfer learning}~\cite{gong2012geodesic} techniques.

Incremental learning extends the number of classes that a deep learning network can support~\cite{xiao2014error}.  Attackers may leverage the technique to empower a model.  Figure~\ref{fig:example_inc} demonstrates the effectiveness of a simple incremental learning approach.  In this experiment, the original model is trained with the CIFAR-10 dataset, and it supports 10 classes.  Our incremental learning program empowers the model to support a new dataset (\textit{i.e.,} CIFAR-100) with 100 classes.  We change the last fully-connected layer of the CNN to support 100 classes, and then train the model with the target dataset.  Finally, the model can achieve an accuracy of 67.95\% after 200 epochs.  In comparison, if we train the same CNN from scratch, the best accuracy is only 58.49\% with the same training settings.

Transfer learning adapts the domain differences between a target dataset and what a model supports~\cite{gong2012geodesic}.  It is an effective machine learning approach when there are insufficient training data in a target dataset.  Attackers may employ the technique to pirate a new model and adapt it to their own scenarios.  Figure~\ref{fig:example_trans} demonstrates the effectiveness of a transfer learning experiment.  In this experiment, the original model is trained with the CIFAR-10 dataset.  The target dataset is STL10~\cite{stl10}, which contains only 500 labeled images for each class.  If we train the model directly with the target dataset~\cite{he2016deep}), we only achieve an accuracy of 58.67\%.  However, if we adopt a straightforward transfer learning technique (\textit{i.e.,} ResNet-TLFT in Figure~\ref{fig:example_trans}), the accuracy can improve to 77.94\%.

In the two examples above, we both fine-tune the original feature extractors to achieve a good accuracy.  Note that fine-tuning can usually help attackers to obtain a better model than without fine-tuning.  For example, in our transfer learning example in Figure~\ref{fig:example_trans}, if we freeze the feature extraction layers and only tune the fully-connected layer (\textit{i.e.,} ResNet-TLFF in Figure~\ref{fig:example_trans}), the model accuracy can only achieve 72.91\%.

\section{Deep Learning Obfuscation}~\label{sec:obfuscation}
\subsection{Definition}
Now we discuss the concept of \textit{deep learning obfuscation}.  It aims to transform the inference logic of a well-trained deep learning model to an obfuscated version, which can prevent attackers from learning its structural design or reusing its well-tuned parameters.  The obfuscated model should retain an equivalent inference function as the original model contains.  Meanwhile, it should incur very limited overhead and demonstrate adequate resilience to attackers.

We borrow the term ``obfuscation'' from classic code obfuscation problems because their purposes are similar.  According to Collberg \textit{et al.}~\cite{collberg1997taxonomy}, classic code obfuscation scrambles code spinets into unintelligible versions while preserving the semantics.  It can increase the difficulties for attackers to interpret the real code logic or to further tamper the code.  

However, deep learning obfuscation is a bit different from classic software obfuscation.  In particular, the parameters of deep learning models are automatically generated via a training process.  To tamper a model, attackers do not need to understand the model, and he can simply fine-tune it with new data.  In comparison, classic software is written by programmers.  Attackers cannot modify it without understanding the details.

\subsection{Performance Metrics}
A competent deep learning obfuscation approach should demonstrate good performance in the following aspects.
\begin{itemize}
\item \textit{Cost}: Cost measures the extra model size and inference time incurred by obfuscation.  A practical obfuscation approach should not incur too much overhead.  Otherwise, it would be useless for real-world applications.

\item \textit{Information leakage}: It shows how much structural information is leaked by the obfuscated model.  The metric corresponds to structure piracy issues.  A potent obfuscation approach should leak as little information as possible.

\item \textit{Fine-tuning ability}: It measures the how much learning ability declination that an obfuscated model suffers.  We can measure the metric by comparing the performance of an original model and an obfuscated model when performing the same fine-tuning task.  It reflects the resilience of an obfuscated model to the parameter piracy attacks.  A network with poor fine-tuning ability can deter attackers from fine-tuning the model parameters with new training data.

\item \textit{Resilience}: The metric reflects the resistance of an obfuscated model to deobfuscation attacks.  In our scenario, deobfuscation means recovering the original structure of the network or empowering an obfuscated model with better fine-tuning ability.

\item \textit{Scalability}: It indicates whether an obfuscation approach can be applied to different deep learning models.  In this work, we claim the effectiveness of our obfuscation approach for convolutional neural network only.
\end{itemize}

Note that resilience evaluates the security of an obfuscation approach, while information leakage and fine-tuning ability evaluate its effectiveness (or potency).  Next, we introduce our structural obfuscation approach for protecting CNN models and then evaluate the performance of our obfuscation approach with respect to these metrics.

\section{Structural Obfuscation Approach}~\label{sec:approach}
For state-of-the-art CNN models, the essential valuable design lies in the structure of their feature extractors.  Our structural obfuscation approach, therefore, aims to hide the structures of the feature extraction networks via simulation.  Below, we first introduce our basic idea of model simulation and then demonstrate how to employ the idea to obfuscate CNN models.

\subsection{Basic Idea}

\begin{figure}[t]
\centering
\includegraphics[width=0.49\textwidth]{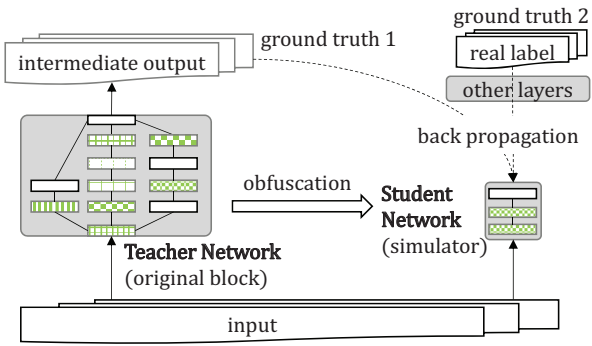}
\caption{Our joint training approach to simulate the feature extractor of a CNN model.}
\label{fig:idea}
\end{figure}

In high level, our obfuscation idea simulates the feature extractor of a CNN model with a shallow and sequential network.  In this way, the simulation network does not reveal the internal structure of the original feature extractor.  Also, by employing a shallow network as the simulator, we intend to lower the fine-tuning ability of the obfuscated model with respect to parameter piracy attacks.

To obtain a competent simulation network, we propose a novel \textit{joint training} idea, which employs two different ground truths to train the simulation network jointly.  Our first ground truth is the output of the target feature extractor, which is the output of a hidden layer.  A CNN model generally outputs more features in a hidden layer than the number of classes in the output layer.  In this way, employing the intermediate output to train a simulation network can provide more fine-grained information than pure labels, and it will facilitate the learning process.  The idea is also known as hint-based training~\cite{romerofitnets}.  In hint-based training, there is a teacher network and a student network, and the student network learns the intermediate knowledge generated by the hidden layers of the teacher network.  However, employing a pure hint-based training approach is insufficient for our scenario because the original hint-based training approach is proposed for model compression scenarios.  It compresses a teacher network into a student network which is deep and thin.  Our approach, on the other hand, aims to simulate a teacher network with a shallow network.  Because a shallow network tends to have worse learning ability, it may not learn the teacher network as well as a deep network.  If there are simulation errors, we hope to mitigate the influence of the error with the information of the real label.  Therefore, we also employ the label of raw input as our second ground truth for training.  Figure~\ref{fig:idea} demonstrates our joint training idea.

Let $\mathcal{T}$ denote the original feature extractor of a teacher network, $\mathcal{S}$ denotes the simulator of a student network, $\mathcal{S'}$ denotes a mixed network that connects the simulator with the rest layers of the original network, and $L_x$ denotes the real labels of the input data $x$.  We employ the following objective function to train the simulator.

\begin{equation}
\argmin_w\left\Vert \mathcal{T}(x) - \mathcal{S}_w(x) + \alpha (L_x - \mathcal{S'}_w(x)) \right\Vert,
\end{equation}
where $ \Vert\cdot\Vert $ represents the $L1$-norm, and $w$ represents the parameters of the simulator.

Because the simulation network is shallow, we may not train it as well as the original model.  In this situation, the real labels can serve as an error corrector or a regulator.  By choosing an appropriate value for $\alpha$, we can balance the resulting impacts to the learning process.

In our later evaluation section, we adopt a tricky approach to search optimal solutions for the joint optimization problem.  We assume the two types of losses are comparable with an appropriate $\alpha$, then we can employ the two ground truths iteratively to train the simulator.

\subsection{Obfuscation Framework}
\begin{figure*}[t]
\centering
\includegraphics[width=0.99\textwidth]{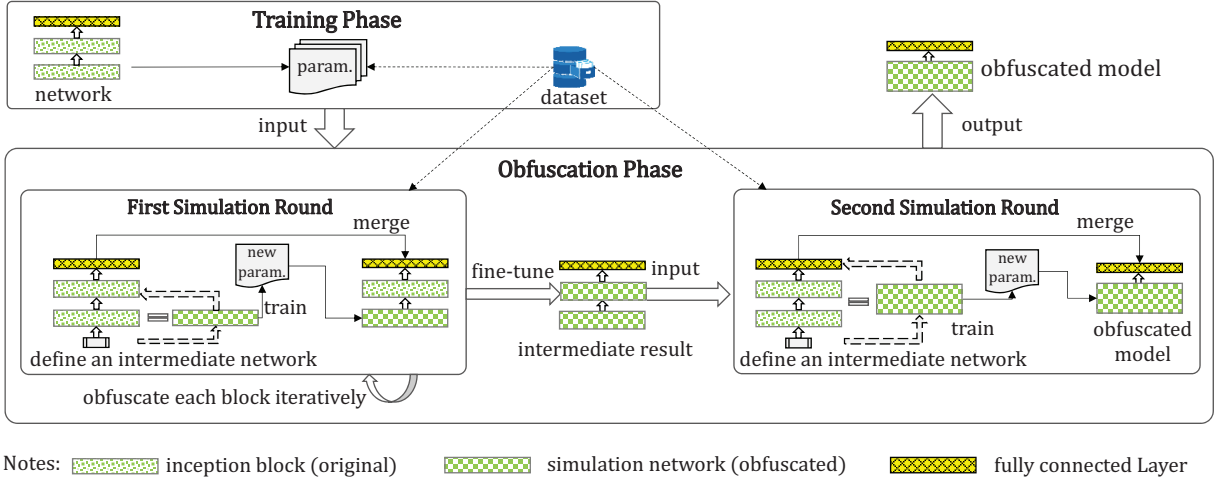}
\caption{Our framework to obfuscate CNN models.}
\label{fig:framework}
\end{figure*}

Because real-world CNNs are very complex, we may not be able to simulate a feature extractor directly.  To achieve an acceptable simulation result, we propose to simulate a feature extractor recursively.  To elaborate, we first simulate each inception block of an extractor with a small simulator.  Then we fine-tune the resulting model to achieve a good accuracy.  In the next round, we simulate the new feature extractor which contains several small simulation networks.  

The first round of simulation obtains two benefits.  On one hand, we can lower the complexity of the target feature extractor to simulate.  On the other hand, it examines whether a target model can be simulated with a shallow network.  If we cannot simulate a feature extractor with a deep network of several small simulators, we are unlikely to be able to simulate it with one shallower network.  After this round, we usually get an intermediate model with an equivalent or slightly higher accuracy.

Our obfuscation framework is demonstrated in Figure~\ref{fig:framework}.  We insert an obfuscation phase after the training phase and before the model is released to clients.  The obfuscation phase contains two rounds.  In the first round, we obfuscate each inception block of the model iteratively.  The obfuscation starts from the bottom layer and goes up to the top layer.  In this order, the residual errors of the lower layers can be mitigated when obfuscating the upper layers.  There are several steps in each iteration.

\begin{enumerate}%[label=(\roman*)]
\item \textit{Define a simulation network:} We first create a small simulation network with a shallow and sequential structure.  Then we plug the network into the original computation graph of the model and make its input and output the same as those of the original inception block.  We defer our discussion about designing the structure of a simulation network to Section~\ref{sec:strategy}.

\item \textit{Train the network}: We tune the parameters of the simulation network with the joint training idea.  The training data is the same as those for training the original model.  To preserve the weights of other layers, we freeze all the parameters before and after the simulator.

\item \textit{Merge the model:} We delete the original inception block from the original network and merge the corresponding parameters.  This step outputs an intermediate obfuscated model.  The obfuscation procedure goes to the next iteration until all the inception blocks have been obfuscated.
\end{enumerate}

When all the inception blocks have been simulated, we fine-tune the model to achieve better accuracy.  Now we can obtain a model $M'$ (intermediate result) with only sequential connections, which is the input to the next simulation round.  In the next round, we also employ the joint training approach to simulate the entire feature extractor of $M'$ at a time.  

Note that the hint-based training approach requires that the teacher network and the student network should have the same dimension of output.  So our intermediate result $M'$ still leaks the information about the interfaces (\textit{e.g.,} feature numbers) between neighboring inception blocks.  Our second round of simulation can further hide such information.  In this way, the finally obfuscated model leaks little information about the internal structure of a feature extractor.

\subsection{Design of Simulation Networks}~\label{sec:strategy}
We should design the simulation networks carefully because they determine the performance of an obfuscated model.  If  the structure of a simulation network is too simple, it may not be able to simulate well.  If it is too complicated, the obfuscated model would incur too much overhead.

When designing a simulation network, our main principle is to maintain the corresponding input and output relationships of the original inception block.  For example, if a feature of output corresponds to a $5 \times 5$ image of input in the original network, we should guarantee that the simulation network also computes the feature based on the $5 \times 5$ image.  To fulfill the principle, we should choose an appropriate combination of convolutional layers and kernel sizes.  Below we elaborate more on the principle. 

\subsubsection{Warm Up}
We first discuss an ideal scenario which assumes no nonlinear operators in an inception block.  In this case, we can simulate the inception block precisely with only one convolutional layer.  We draw this conjecture based on Rule~\ref{rule:sequential} and Rule~\ref{rule:parallel}.

\begin{ruler}\label{rule:sequential}
For any two-layered sequential convolutional block with no nonlinear operators, there exists an equivalent convolutional layer to simulate it.
\end{ruler}

\begin{proof}
To show that we can simulate any two-layered convolutional layers with only one layer, we show that the feature images outputted by a two-layered convolutional layers are linear to the input.

Supposing the first convolutional layer inputs images $X$ of $c_{in}$ channels and outputs feature images $Y$ of $c_{l1}$ channels, we can compute a pixel value $y$ of $Y$ as

\begin{equation}\label{eq:rule1}
y=\sum_{i=1}^{c_{in}} f(K, X_i) + \beta,
\end{equation}
where $K$ is a kernel matrix, $X_i$ is a corresponding image of the $i$th channel, and $f(K, X_i) = k_{1,1}x_{1,1}+k_{1,2}x_{1,2}+...+k_{h_{l1},w_{l1}}x_{h_{l1},w_{l1}}$, where $h_{l1} \times w_{l1}$ is the kernel size of the first convolutional layer.  

Because $K$ and $\beta$ are constants in a pre-trained model and only $X_i$ contains variables, we can simplify Equation~\ref{eq:rule1} as 

\begin{equation}
y = \alpha_1 x_{1,1,1}+...+\alpha_{c_{l1},h_{l1},w_{l1}} x_{c_{l1},h_{l1},w_{l1}},
\end{equation}
where $\alpha_i$ is a constant, and $y$ is linear to $X$.  

Similarly, supposing the second convolutional layer outputs feature images $Z$ of $c_{l2}$ channels, we can compute a pixel value $z$ of $Z$ with the form of $z = \alpha_1 y_{1,1,1}+...+\alpha_{c_{l2},h_{l2},w_{l2}}y_{c_{l1},h_{l1},w_{l1}}$.  Unfolding each $y$, we can get 

\begin{equation}
z = \lambda_1 x_{1,1,1}+...+\lambda_{q} x_{c_{in},h,w}, 
\end{equation}
where $\lambda_i$ is a constant, $h = h_{l1}+(h_{l2}-1)s$, $w = w_{l1}+(w_{l2}-1)s$, and $s$ is the stride of the first convolutional layer.
\end{proof}

\begin{ruler}\label{rule:parallel}
For any two-paralleled convolutional layers merged with concatenation or add, there exists an equivalent convolutional layer to simulate it.
\end{ruler}

\begin{proof}

Suppose the input images are $X$ with $c_{in}$ channels, the first convolutional layer outputs features $Y$ of $c_{1}$ channels, and the second convolutional layer outputs features $Z$ of $c_{2}$ channels.  A pixel value $y$ of $Y$ can be computed as

\begin{equation}
y=\sum_{i=1}^{c_{in}} f(K_{y}, X_i) + \beta,
\end{equation}
and a pixel value $z$ of $Z$ can be computed as

\begin{equation}
z=\sum_{i=1}^{c_{in}} f(K_{z}, X_i) + \beta,
\end{equation}
where $K_{y}$ is a kernel of the first convolutional layer, and $K_{z}$ is a kernel of the second convolutional layer.

If the layers are merged with concatenation, we can directly substitute the block with an equivalent convolutional layer with a channel size $c_{1}+ c_{2}$ and a kernel size $Max(Size(K_{y}),Size(K_{z}))$.  If the layers are merged with add, the two convolutional layers should have an equivalent number of channels, $c_1$ and $c_2$.
\end{proof}

Given any inception block without nonlinear operators, we can apply Rule~\ref{rule:sequential} and Rule~\ref{rule:parallel} iteratively to compress the module into one convolutional layer.  The channel number of the resulting convolutional layer should be equivalent to the original inception block.  Its kernel size should be selected as the max kernel size when simulating each convolutional sequence.

\subsubsection{Choose an Appropriate Structure}
When an inception block has nonlinear operators, we cannot simulate the module precisely.  However, we can still apply Rule~\ref{rule:sequential} and Rule~\ref{rule:parallel} to compute an appropriate kernel size.  Take the inception block of ResNet (Figure~\ref{fig:resnet_incep}) as an example, we can compute that each feature corresponds to a $5\times5$ image, so that we can simulate the block with one convolutional layer whose kernel size is 5, or with two convolutional layers whose kernel sizes are both 3.  If an inception block has pooling operations (\textit{e.g.,} the inception block of GoogLeNet (Figure~\ref{fig:googlenet_incep})), we treat them as same as convolution, because they are also kernel-based operations.

In general, we can design a simulation network with two to four convolutional layers to simulate an inception block.  By default, we apply both batch normalization and ReLU after each convolutional layer.  The specific choice generally depends on the complexity of the inception block to simulate.  For example, to simulate one inception block of GoogleNet, two convolutional layers are sufficient.  However, if the target block has tens of convolutional layers, we may employ a simulation network with more convolutional layers.  We will discuss more details about how we design the simulation network for several practical deep learning models in the evaluation section (Section~\ref{sec:evaluation}). 

\begin{figure*}[t]
\centering
\subfigure[Obfuscating GoogLeNet.]{
\label{fig:googlenet}
\includegraphics[width=0.75\textwidth]{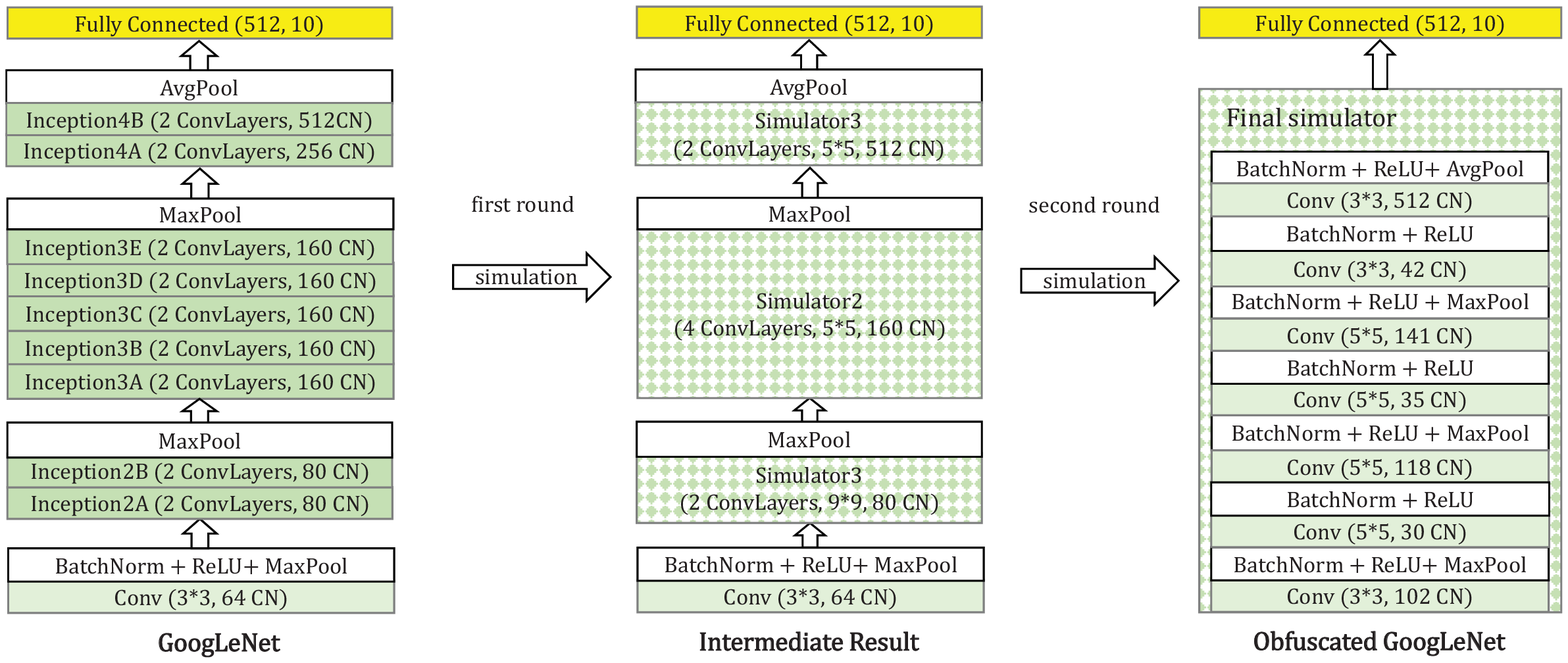}}
\subfigure[Obfuscating ResNet.]{
\label{fig:resnet}
\includegraphics[width=0.75\textwidth]{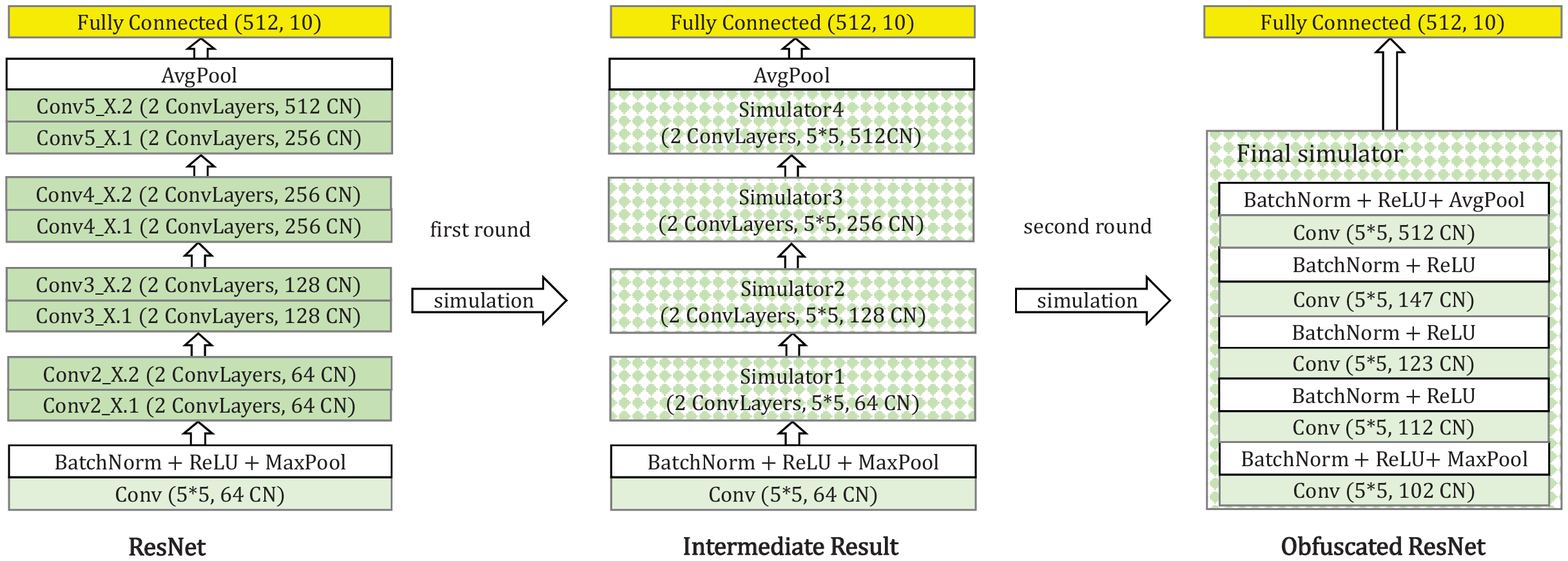}} 
\subfigure[Obfuscating DenseNet.]{
\label{fig:densenet}
\includegraphics[width=0.75\textwidth]{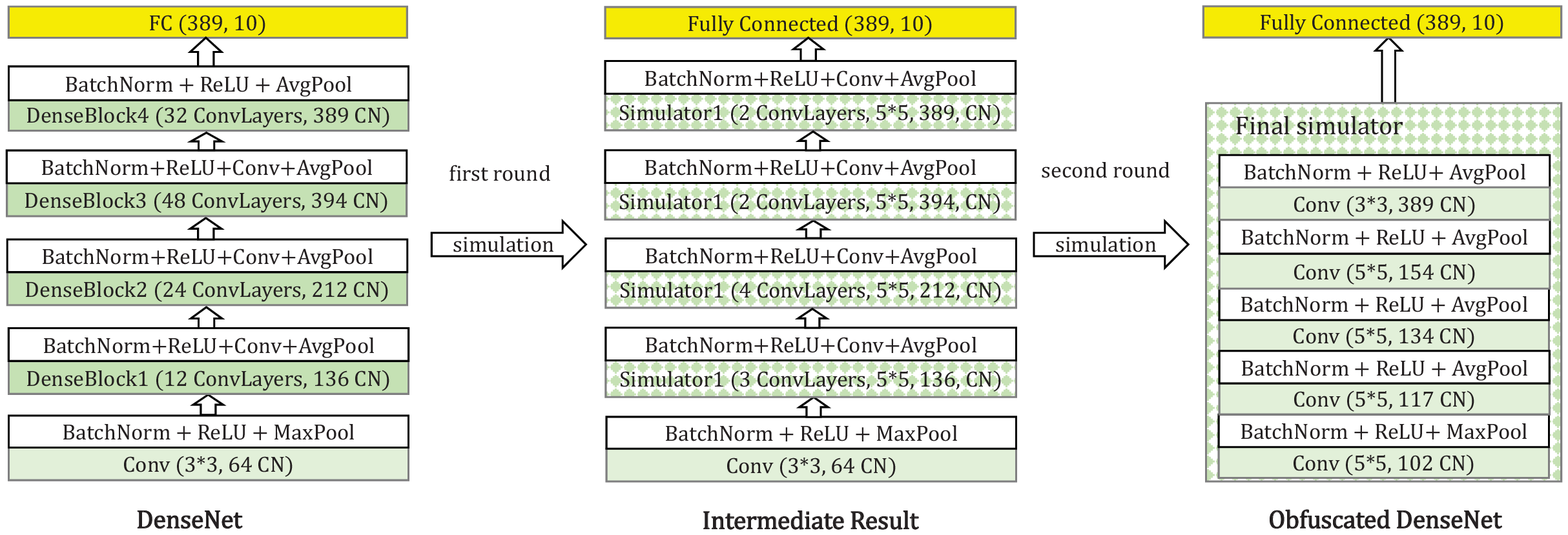}}
\caption{The structures of CNN models and their simulation networks employed in our experiments. CN: channels.}
\label{fig:networks}
\end{figure*}

\section{Evaluation}~\label{sec:evaluation}
In this section, we examine the performance of our obfuscation approach with real-world experiments.  We focus on three aspects: 1) whether an obfuscated model can achieve negligible accuracy declination in comparison with the original model; 2) how much overhead will be incurred by our obfuscation approach; 3) and whether the fine-tuning abilities of obfuscated models become worse.
  
\subsection{Experimental Setting}
To show that our obfuscation approach has no bias on particular CNN structures, we choose three prevalent CNNs to obfuscate, \textit{i.e.,} GoogLeNet~\cite{szegedy2015going}, ResNet~\cite{he2016deep} and DenseNet~\cite{huang2017densely}.  Figure~\ref{fig:networks} demonstrates the architecture of the networks we adopted in our experiments, and we train them with the CIFAR-10 dataset.  To fit the CNN structures for the dataset, we have slightly changed their original settings. 

The CIFAR-10 dataset~\cite{cifar10} contains 10 classes of $32\times32$ colorful images (\textit{e.g.,} airplane, automobile, bird).  Each class contains 5000 images for training and 1000 images for testing.  The dataset is widely employed to benchmark the performance of CNNs, such as ResNet~\cite{he2016deep} and DenseNet~\cite{huang2017densely}.  We train the CNN models using the suggested settings of learning rates until the testing accuracy cannot get further improved.  In our experiment, it takes one thousand epochs each to train a model.  Our GoogLeNet model finally reaches an accuracy of 90.83\%, ResNet reaches 90.94\%, and DenseNet reaches 90.14\%.  The detailed training results are demonstrated in Figure~\ref{fig:googlenet_train}, Figure~\ref{fig:resnet_train}, and Figure~\ref{fig:densenet_train} respectively.

To conduct obfuscation experiments, we write all of our deep learning programs with Pytorch 3.1 and experimental scripts with Python 3.6.  Our experimental host is an Ubuntu (version 16.04) server with a Xeon CPU (E5-2650) and 128G memory, and the experimental GPU is Nvidia Titan Xp Pascal with 12G memory.  

\subsection{Steps of Obfuscation} 
\subsubsection{Simulating GoogLeNet} 
Figure~\ref{fig:googlenet} demonstrates our obfuscation steps for GoogLeNet.  The original GoogLeNet has nine inception blocks, each of which is demonstrated as Figure~\ref{fig:googlenet_incep}.  These nine inception blocks are separated into three groups by max pooling operations: $\{2A,2B\}$, $\{3A,...,3E\}$, and $\{4A,4B\}$.  

We finally obfuscate GoogLeNet with a five-layered sequential block.  To this end, we adopt the recursive simulation idea.  In the first simulation round, we simulate each group of inception blocks with a sequential block.  For the groups of $\{2A,2B\}$ and $\{4A, 4B\}$, we employ a two-layered convolutional block each.  For the group of $\{3A,...,3E\}$, we employ a four-layered convolutional block, and for $\{4A, 4B\}$ with a three-layered convolutional block.  Then we fine-tune the resulting model to obtain an intermediate result.  Note that when simulating each group of inception blocks, we freeze all the parameters of other layers.  When fine-tuning the model, we tune all the parameters of the intermediate model.  Then in the second simulation round, we employ the fine-tuned intermediate model to train a six-layered convolutional block, which is the feature extractor of our final obfuscated GoogLeNet.

\subsubsection{Simulating ResNet} 
Figure~\ref{fig:resnet} demonstrates our process to obfuscate ResNet.  We employ ResNet-18, which contains four inception blocks, and each block has two sub-blocks.  In the first round, we simulate each inception block of the network at one time with a two-layered convolutional block.  In the second round, we simulate the whole feature extractor of the intermediate model with a five-layered convolutional block.

\subsubsection{Simulating DenseNet} 
Figure~\ref{fig:densenet} shows how we obfuscate DenseNet-121.  It has four dense blocks, and each block has 12, 24, 48, or 32 convolutional layers respectively.  Their connection mode is demonstrated as Figure~\ref{fig:densenet_incep}.  In the first round, we simulate each dense block at one time.  The four simulation networks have 3,4,2,2 convolutional layers respectively.  In the second round, we simulate the whole feature extractor at one time with a five-layered convolutional block.

\subsection{Performance of Obfuscated Models}

\begin{table}[t]
\footnotesize
\centering
\caption{Evaluation results about the performance of our obfuscated models.  For each model, we report the accuracy, model size, and average inference time.  The overhead is computed as $\frac{cost_{2}}{cost_{1}}-1$.}
\label{tab:results}
\newcommand{\tabincell}[2]{\begin{tabular}{@{}#1@{}}#2\end{tabular}}
\renewcommand{\multirowsetup}{\centering} 
\begin{tabular}{|c|c|c|c|c|c|c|}
\hline
\multirow{2}{*}{\textbf{\tabincell{c}{Model}}} & \multirow{2}{*}{\textbf{\tabincell{c}{Sim.\\ Round}}} & \multicolumn{3}{|c|}{\textbf{Performance}}  & \multicolumn{2}{|c|}{\textbf{Overhead}} \\
\cline{3-7} 
& & acc. & \tabincell{c}{size\\(MB)} & \tabincell{c}{time\\($\mu s$)} & size & time \\
\hline
\multirow{3}{*}{\tabincell{c}{GoogLe\\-Net}} & - & 90.83\% & 2.51 & 17.85 & - & - \\
\cline{2-7}
& 1st & 90.99\% & 7.82 & 7.69 & 212\% & -59\%  \\
\cline{2-7}
& 2nd & 90.92\% & 2.49 & 7.01 & -1\% & -63\%  \\
\hline
\multirow{3}{*}{\tabincell{c}{ResNet}} & - & 90.94\% & 43.36 & 10.50 & - & - \\
\cline{2-7}
& 1st & 91.39\% & 26.80 & 6.76 & -38\% & -36\%\\
\cline{2-7}
& 2nd & 91.04\% & 11.38 & 5.17 & -74\% & -51\%\\
\hline
\multirow{3}{*}{\tabincell{c}{Dense\\-Net}} & - & 90.14\% & 4.24 & 35.53 & - & - \\
\cline{2-7}
& 1st & 90.87\% & 8.86 & 8.86 & 109\% & -75\%\\
\cline{2-7}
& 2nd & 90.31\% & 4.21 & 5.52 & -1\% & -84\% \\
\hline
\end{tabular}
\end{table}

Table~\ref{tab:results} demonstrates our experimental results when obfuscating the three models.  For each obfuscation experiment, we report the performance of the original model, the intermediate model achieved after the first round of simulation, and the finally obfuscated model.  For each of the models, we measure its performance with testing accuracy, model size, and average inference time.  From the results, we can observe that all our finally obfuscated models suffer no accuracy declination, and some are even more efficient than the original models.  

For ResNet, our finally obfuscated model saves $74\%$ size of the original model, and it is about two times faster than the original model.  This is not very surprising, because ResNet is not very efficient in model size.  Note that the inception block of ResNet adds up the results of two parallel convolutional sequences as its output.  Therefore, it needs more parameters than a sequential network to compute the same number of features.  

The results show that we can also obfuscate GoogLeNet and DenseNet with no size overhead, and the obfuscated models can save 63\% and 84\% average inference time correspondingly.  This is not easy because the two networks are already very efficient in model size.  Their inception blocks concatenate the resulting feature images of parallel convolutional sequences.  In this way, they need fewer parameters than a sequential network to compute the same number of features.  Although our intermediate results achieved in the first round incur extra size overhead, such overhead can be fully mitigated after the second round.

Figure~\ref{fig:obfuscation_details} demonstrates more detailed records about each round of our obfuscation experiments.  From the figures, we can observe that after the first simulation round, the resulting model generally suffers some accuracy declination.  However, we can fine-tune the intermediate model (with the data labels as the ground truth) to achieve better accuracy.  This is not because the architecture of the intermediate model is more powerful, but because it employs a good initial state for fine-tuning.  Note that we can hardly improve the accuracy of the original models before.  The interesting phenomenon provides some benefits for our second round of simulation.  The recursive simulation, therefore, can help us to approach a more competent obfuscation result with better accuracy.  In some cases, if we could not obtain an intermediate model with slightly better or at least equivalent accuracy, it implies the simulation network may not be powerful enough, and we should empower the simulation network, such as by adjusting the number of convolutional layers or channels.

\begin{figure*}[t]
\centering

\subfigure[GoogLeNet training process.]{
\label{fig:googlenet_train}
\begin{tikzpicture}[scale=0.65]
\begin{axis}[xlabel=epoch, ylabel=accuracy, legend pos=south east, height=5cm]

\addplot[red, dashed, thick] table[mark=none, x=Epoch,y=TrainAcc] {Experiment/googlenet/train.txt};
\addplot[blue, solid, thick] table[mark=none, x=Epoch,y=TestAcc] {Experiment/googlenet/train.txt};

\legend{TrainAcc, TestAcc}
\end{axis}
\end{tikzpicture}
}
\subfigure[The 1st round of GoogLeNet simulation.]{
\label{fig:googlenet_sim1}
\begin{tikzpicture}[scale=0.65]
\begin{axis}[xlabel=epoch, ylabel=accuracy, legend pos=south east, height=5cm]

\addplot[red, solid, thick] table[mark=none, x=Epochs,y=Similator1] {Experiment/googlenet/1st_round.txt};
\addplot[blue, dashed, thick] table[mark=none, x=Epochs,y=Similator2] {Experiment/googlenet/1st_round.txt};
\addplot[green, dotted, thick] table[mark=none, x=Epochs,y=Similator3] {Experiment/googlenet/1st_round.txt};

\legend{Simulator1, Simulator2, Simulator3}
\end{axis}
\end{tikzpicture}
}
\subfigure[Fine-tune the intermediate GoogLeNet.]{
\label{fig:googlenet_finetune}
\begin{tikzpicture}[scale=0.65]
\begin{axis}[xlabel=epoch, ylabel=accuracy, legend pos=south east, height=5cm]

\addplot[red, dashed, thick] table[mark=none, x=Epoch,y=TrainAcc] {Experiment/googlenet/finetune.txt};
\addplot[blue, solid, thick] table[mark=none, x=Epoch,y=TestAcc] {Experiment/googlenet/finetune.txt};

\legend{TrainAcc, TestAcc}
\end{axis}
\end{tikzpicture}
}
\subfigure[The 2nd round of GoogLeNet simulation.]{
\label{fig:googlenet_sim2}	
\begin{tikzpicture}[scale=0.65]
\begin{axis}[xlabel=epoch, ylabel=accuracy, legend pos=south east, height=5cm]

\addplot[red, dashed, thick] table[mark=none, x=Epoch,y=TrainAcc] {Experiment/googlenet/2nd_round.txt};
\addplot[blue, solid, thick] table[mark=none, x=Epoch,y=TestAcc] {Experiment/googlenet/2nd_round.txt};

\legend{TrainAcc, TestAcc}
\end{axis}
\end{tikzpicture}
}
\subfigure[ResNet training process.]{
\label{fig:resnet_train}	
\begin{tikzpicture}[scale=0.65]
\begin{axis}[xlabel=epoch, ylabel=accuracy, legend pos=south east, height=5cm]

\addplot[red, dashed, thick] table[mark=none, x=Epoch,y=TrainAcc] {Experiment/resnet/train.txt};
\addplot[blue, solid, thick] table[mark=none, x=Epoch,y=TestAcc] {Experiment/resnet/train.txt};

\legend{TrainAcc, TestAcc}
\end{axis}
\end{tikzpicture}
}
\subfigure[The 1st round of ResNet simulation.]{
\label{fig:resnet_sim1}	
\begin{tikzpicture}[scale=0.65]
\begin{axis}[xlabel=epoch, ylabel=accuracy, legend pos=south east, height=5cm]

\addplot[red, solid, thick] table[mark=none, x=Epochs,y=Similator1] {Experiment/resnet/1st_round.txt};
\addplot[blue, dashed, thick] table[mark=none, x=Epochs,y=Similator2] {Experiment/resnet/1st_round.txt};
\addplot[green, dotted, thick] table[mark=none, x=Epochs,y=Similator3] {Experiment/resnet/1st_round.txt};
\addplot[brown, thick] table[mark=none, x=Epochs,y=Similator4] {Experiment/resnet/1st_round.txt};

\legend{Simulator1, Simulator2, Simulator3, Simulator4}
\end{axis}
\end{tikzpicture}
}
\subfigure[Fine-tune the intermediate ResNet.]{
\label{fig:resnet_finetune}	
\begin{tikzpicture}[scale=0.65]
\begin{axis}[xlabel=epoch, ylabel=accuracy, legend pos=south east, height=5cm]

\addplot[red, dashed, thick] table[mark=none, x=Epoch,y=TrainAcc] {Experiment/resnet/finetune.txt};
\addplot[blue, solid, thick] table[mark=none, x=Epoch,y=TestAcc] {Experiment/resnet/finetune.txt};

\legend{TrainAcc, TestAcc}
\end{axis}
\end{tikzpicture}
}
\subfigure[The 2nd round of ResNet simulation.]{
\label{fig:resnet_sim2}	
\begin{tikzpicture}[scale=0.65]
\begin{axis}[xlabel=epoch, ylabel=accuracy, legend pos=south east, height=5cm]

\addplot[red, dashed, thick] table[mark=none, x=Epoch,y=TrainAcc] {Experiment/resnet/2nd_round.txt};
\addplot[blue, solid, thick] table[mark=none, x=Epoch,y=TestAcc] {Experiment/resnet/2nd_round.txt};

\legend{TrainAcc, TestAcc}
\end{axis}
\end{tikzpicture}
}

\subfigure[DenseNet training process.]{
\label{fig:densenet_train}
\begin{tikzpicture}[scale=0.65]
\begin{axis}[xlabel=epoch, ylabel=accuracy, legend pos=south east, height=5cm]

\addplot[red, dashed, thick] table[mark=none, x=Epoch,y=TrainAcc] {Experiment/densenet/train.txt};
\addplot[blue, solid, thick] table[mark=none, x=Epoch,y=TestAcc] {Experiment/densenet/train.txt};

\legend{TrainAcc, TestAcc}
\end{axis}
\end{tikzpicture}
}
\subfigure[The 1st round of DenseNet simulation.]{
\label{fig:densenet_sim1}
\begin{tikzpicture}[scale=0.65]
\begin{axis}[xlabel=epoch, ylabel=accuracy, legend pos=south east, height=5cm]

\addplot[red, solid, thick] table[mark=none, x=Epochs,y=Similator1] {Experiment/densenet/1st_round.txt};
\addplot[blue, dashed, thick] table[mark=none, x=Epochs,y=Similator2] {Experiment/densenet/1st_round.txt};
\addplot[green, dotted, thick] table[mark=none, x=Epochs,y=Similator3] {Experiment/densenet/1st_round.txt};
\addplot[brown, thick] table[mark=none, x=Epochs,y=Similator4] {Experiment/densenet/1st_round.txt};

\legend{Simulator1, Simulator2, Simulator3, Simulator4}
\end{axis}
\end{tikzpicture}
}
\subfigure[Fine-tune the intermediate DenseNet.]{
\label{fig:densenet_finetune}
\begin{tikzpicture}[scale=0.65]
\begin{axis}[xlabel=epoch, ylabel=accuracy, legend pos=south east, height=5cm]

\addplot[red, dashed, thick] table[mark=none, x=Epoch,y=TrainAcc] {Experiment/densenet/finetune.txt};
\addplot[blue, solid, thick] table[mark=none, x=Epoch,y=TestAcc] {Experiment/densenet/finetune.txt};

\legend{TrainAcc, TestAcc}
\end{axis}
\end{tikzpicture}
}
\subfigure[The 2nd round of DenseNet simulation.]{
\label{fig:densenet_sim2}
\begin{tikzpicture}[scale=0.65]
\begin{axis}[xlabel=epoch, ylabel=accuracy, legend pos=south east, height=5cm]

\addplot[red, dashed, thick] table[mark=none, x=Epoch,y=TrainAcc] {Experiment/densenet/2nd_round.txt};
\addplot[blue, solid, thick] table[mark=none, x=Epoch,y=TestAcc] {Experiment/densenet/2nd_round.txt};

\legend{TrainAcc, TestAcc}
\end{axis}
\end{tikzpicture}
}
\caption{Our detailed experimental results when obfuscating the models of GoogLeNet, ResNet, and DenseNet.  For each model obfuscation experiment, we show the accuracy when training the original model, conducting the first round of simulation, performing fine-tuning, and conducting the second round of simulation.}
\label{fig:obfuscation_details}
\end{figure*}
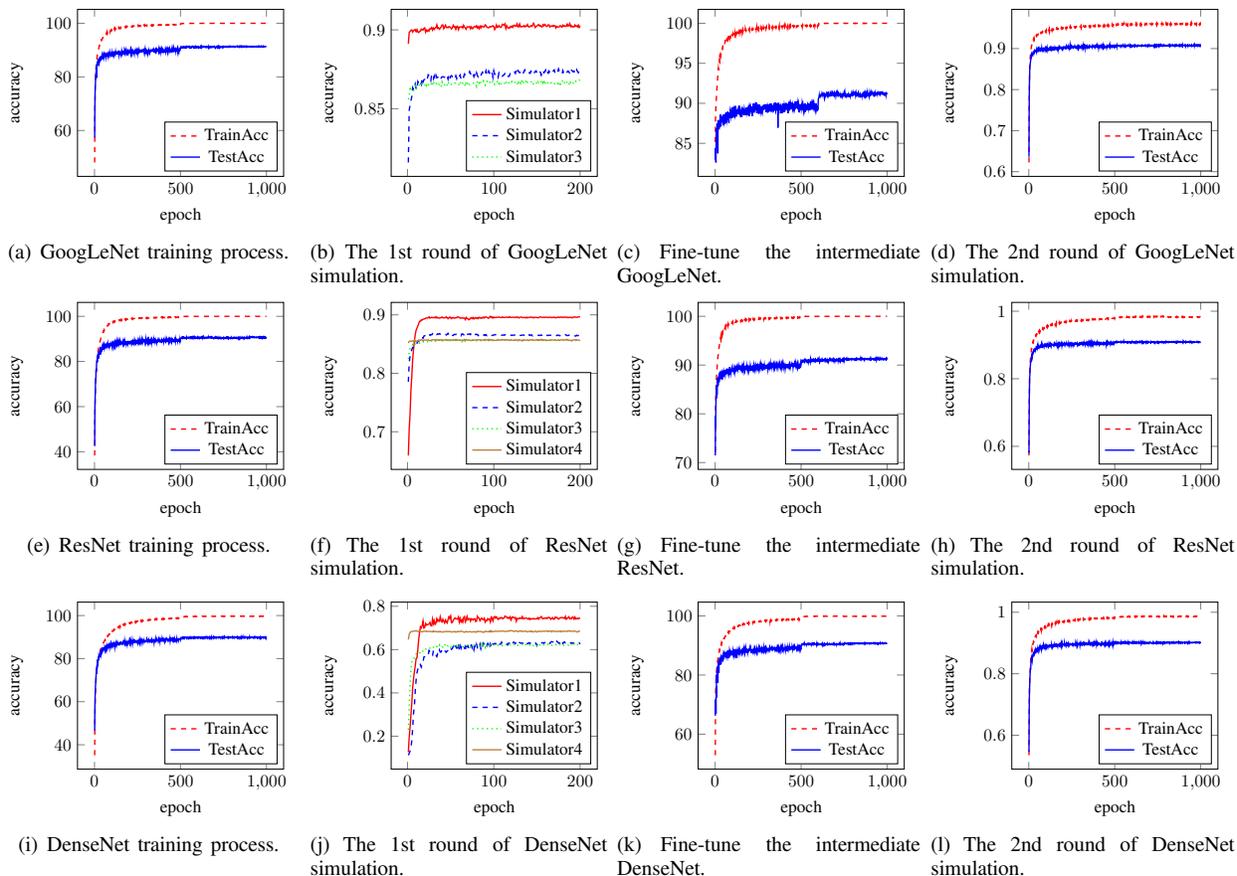

\begin{figure*}[t]
\centering
\subfigure[Incremental learning experiment with GoogLeNet.]{
\label{fig:cifar100_googlenet}
\begin{tikzpicture}[scale=0.65]
\begin{axis}[xlabel=epoch, ylabel=accuracy, legend pos=south east, height=6.5cm]

\addplot[red, dashed] table[mark=none, x=Epoch,y=TrainAcc] {Experiment/googlenet/incremental.txt};
\addplot[red, solid] table[mark=none, x=Epoch,y=TestAcc] {Experiment/googlenet/incremental.txt};
\addplot[blue, dashed] table[mark=none, x=Epoch,y=ObfTrainAcc] {Experiment/googlenet/incremental.txt};
\addplot[blue, solid] table[mark=none, x=Epoch,y=ObfTestAcc] {Experiment/googlenet/incremental.txt};

\legend{GoogLeNet TrainAcc,GoogLeNet TestAcc, Obf. GoogLeNet TrainAcc, Obf. GoogLeNet TestAcc}
\end{axis}
\end{tikzpicture}
}
%\hspace{0.3cm}
\subfigure[Incremental learning experiment with ResNet.]{
\label{fig:cifar100_resnet}
\begin{tikzpicture}[scale=0.65]
\begin{axis}[xlabel=epoch, ylabel=accuracy, legend pos=south east, height=6.5cm]

\addplot[red, dashed] table[mark=none, x=Epoch,y=TrainAcc] {Experiment/resnet/incremental.txt};
\addplot[red, solid] table[mark=none, x=Epoch,y=TestAcc] {Experiment/resnet/incremental.txt};
\addplot[blue, dashed] table[mark=none, x=Epoch,y=ObfTrainAcc] {Experiment/resnet/incremental.txt};
\addplot[blue, solid] table[mark=none, x=Epoch,y=ObfTestAcc] {Experiment/resnet/incremental.txt};

\legend{RestNet TrainAcc,RestNet TestAcc, Obf. RestNet TrainAcc, Obf. RestNet TestAcc}
\end{axis}
\end{tikzpicture}
}
%\hspace{0.3cm}
\subfigure[Incremental learning experiment with DenseNet.]{
\label{fig:cifar100_densenet}
\begin{tikzpicture}[scale=0.65]
\begin{axis}[xlabel=epoch, ylabel=accuracy, legend pos=south east, height=6.5cm]

\addplot[red, dashed] table[mark=none, x=Epoch,y=TrainAcc] {Experiment/densenet/incremental.txt};
\addplot[red, solid] table[mark=none, x=Epoch,y=TestAcc] {Experiment/densenet/incremental.txt};
\addplot[blue, dashed] table[mark=none, x=Epoch,y=ObfTrainAcc] {Experiment/densenet/incremental.txt};
\addplot[blue, solid] table[mark=none, x=Epoch,y=ObfTestAcc] {Experiment/densenet/incremental.txt};

\legend{DenseNet TrainAcc,DenseNet TestAcc, Obf. DenseNet TrainAcc, Obf. DenseNet TestAcc}
\end{axis}
\end{tikzpicture}
}

\subfigure[Transfer learning experiment with GoogLeNet.]{
\label{fig:stl10_googlenet}
\begin{tikzpicture}[scale=0.65]
\begin{axis}[xlabel=epoch, ylabel=accuracy, legend pos=south east, height=6.5cm]

\addplot[red, dashed] table[mark=none, x=Epoch,y=TrainAcc] {Experiment/googlenet/transfer.txt};
\addplot[red, solid] table[mark=none, x=Epoch,y=TestAcc] {Experiment/googlenet/transfer.txt};
\addplot[blue, dashed] table[mark=none, x=Epoch,y=ObfTrainAcc] {Experiment/googlenet/transfer.txt};
\addplot[blue, solid] table[mark=none, x=Epoch,y=ObfTestAcc] {Experiment/googlenet/transfer.txt};

\legend{GoogLeNet TrainAcc,GoogLeNet TestAcc, Obf. GoogLeNet TrainAcc, Obf. GoogLeNet TestAcc}
\end{axis}
\end{tikzpicture}
}
\subfigure[Transfer learning experiment with ResNet.]{
\label{fig:stl10_resnet}
\begin{tikzpicture}[scale=0.65]
\begin{axis}[xlabel=epoch, ylabel=accuracy, legend pos=south east, height=6.5cm]

\addplot[red, dashed] table[mark=none, x=Epoch,y=TrainAcc] {Experiment/resnet/transfer.txt};
\addplot[red, solid] table[mark=none, x=Epoch,y=TestAcc] {Experiment/resnet/transfer.txt};
\addplot[blue, dashed] table[mark=none, x=Epoch,y=ObfTrainAcc] {Experiment/resnet/transfer.txt};
\addplot[blue, solid] table[mark=none, x=Epoch,y=ObfTestAcc] {Experiment/resnet/transfer.txt};

\legend{RestNet TrainAcc,RestNet TestAcc, Obf. RestNet TrainAcc, Obf. RestNet TestAcc}
\end{axis}
\end{tikzpicture}
}
\subfigure[Transfer learning experiment with DenseNet.]{
\label{fig:stl10_densenet}
\begin{tikzpicture}[scale=0.65]
\begin{axis}[xlabel=epoch, ylabel=accuracy, legend pos=south east, height=6.5cm]

\addplot[red, dashed] table[mark=none, x=Epoch,y=TrainAcc] {Experiment/densenet/transfer.txt};
\addplot[red, solid] table[mark=none, x=Epoch,y=TestAcc] {Experiment/densenet/transfer.txt};
\addplot[blue, dashed] table[mark=none, x=Epoch,y=ObfTrainAcc] {Experiment/densenet/transfer.txt};
\addplot[blue, solid] table[mark=none, x=Epoch,y=ObfTestAcc] {Experiment/densenet/transfer.txt};

\legend{DenseNet TrainAcc,DenseNet TestAcc, Obf. DenseNet TrainAcc, Obf. DenseNet TestAcc}
\end{axis}
\end{tikzpicture}
}
\caption{Our detailed experimental results when comparing the fine-tuning abilities of the obfuscated models and their original models.  We employ the STL10 dataset for transfer learning experiments, and the CIFAR-100 dataset for incremental learning experiments.}
\label{fig:finetune}
\end{figure*}
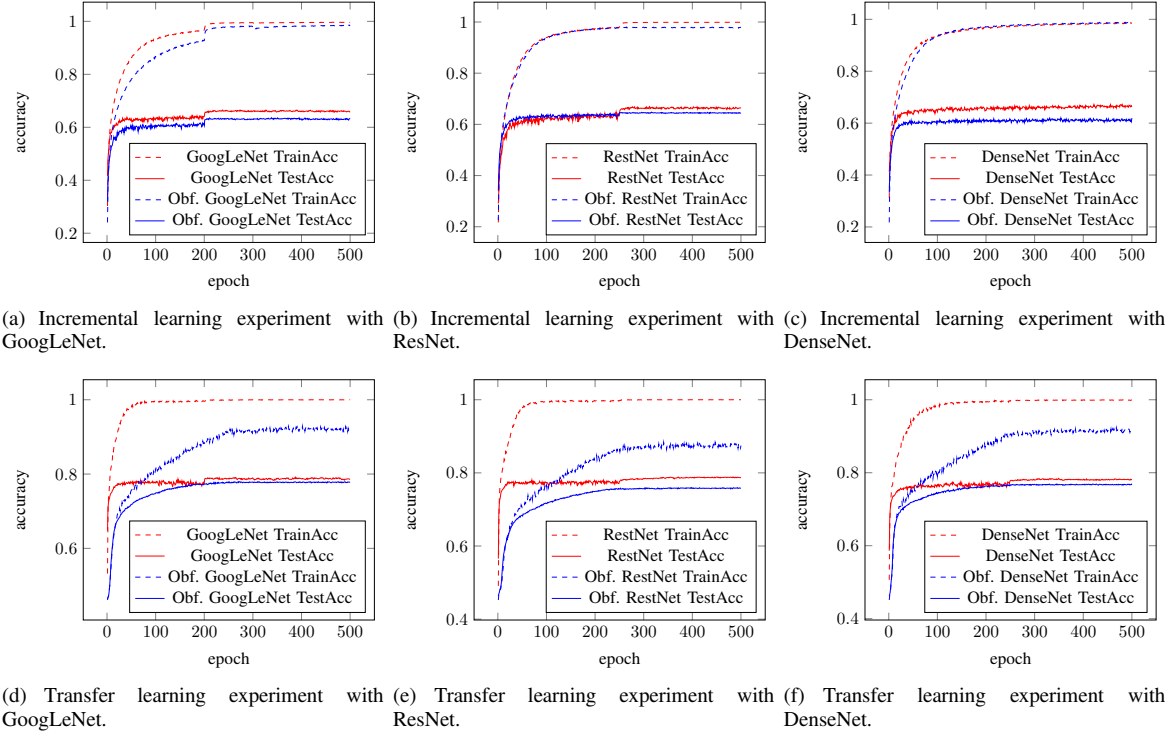

\subsection{Fine-tuning Ability}

\begin{table}[t]
\footnotesize
\centering
\caption{Evaluation results which compare the fine-tuning ability of an obfuscated model and the original model.  For each experiment, we report the accuracy and the declination of accuracy.  The declination is computed as $1-\frac{accuracy_2}{accuracy_1}$.}
\label{tab:finetune}
\newcommand{\tabincell}[2]{\begin{tabular}{@{}#1@{}}#2\end{tabular}}
\renewcommand{\multirowsetup}{\centering} 
\begin{tabular}{|c|c|c|c|c|}
\hline
\multirow{2}{*}{\textbf{\tabincell{c}{Network}}} & \multicolumn{2}{|c|}{\textbf{CIFAR-100}} & \multicolumn{2}{|c|}{\textbf{STL10}}  \\
\cline{2-5} 
& accuracy & decline & accuracy & decline \\
\hline
\tabincell{c}{GoogLeNet}  & 66.5\% & - & 79.15\%& - \\
\hline
\tabincell{c}{Obf. GoogLeNet}  & 63.59\% & 4.4\% & 77.95\% & 1.5\%  \\
\hline
\tabincell{c}{ResNet} & 66.92\% & - & 78.86\% & - \\
\hline
\tabincell{c}{Obf. ResNet} & 64.77\% & 3.2\% & 75.97\% & 3.7\%  \\
\hline
\tabincell{c}{DenseNet} & 67.16\% & - & 78.45\% & - \\
\hline
\tabincell{c}{Obf. DenseNet} & 62.91\% & 6.3\% & 76.90\% & 2.0\%  \\
\hline
\end{tabular}
\end{table}

To evaluate the fine-tuning ability of each obfuscated model, we conduct an incremental learning experiment with the CIFAR-100 dataset~\cite{cifar10}, and a transfer learning experiment with the STL10 dataset~\cite{stl10}.  The CIFAR-100 dataset has 100 classes of images, and each class contains 500 images for training and 100 images for testing.  The STL10 dataset has 10 classes of images, which is the same as CIFAR-10.  However, the original image format of STL10 is $96 \times 96$, and we have to resize the images to $32\times32$ to fit our model.  Besides, there are also only 500 images for training in each class.  Because the training dataset of CIFAR-100 and STL10 are both 10 times smaller than CIFAR-10 for each class, we may not achieve an accuracy as good as CIFAR-10.  Employing transfer learning and incremental learning techniques should be helpful to achieve better accuracy.

Table~\ref{tab:finetune} demonstrates our experimental results.  As a comparison, we also report the corresponding performance of the original model.  Overall, the obfuscated models suffer obvious accuracy declination when performing both the incremental learning and transfer learning tasks, ranging from 1.5\% to 6.3\%.  Our detailed experimental records are demonstrated in Figure~\ref{fig:finetune}.  To compare the performance fair, we employ the same training strategy for each model and its obfuscated model.  The experimental results verify our idea that an obfuscated model with shallow structure should have worse fine-tuning ability than its original model.  

In particular, the declination degrees are related to specific fine-tuning tasks and the structures of simulation networks.  In our experiment, the declinations of the incremental learning experiments are more obvious than those of the transfer learning experiments.  We think one reason is that the incremental learning task is more difficult than transfer learning task because it requires distinguishing $10 \times$ classes.  The fine-tuning ability declination may further exaggerate when pursuing more difficult tasks.  Therefore, the metric of fine-tuning ability is related to particular tasks, and such results only provide developers a relative measurement about the resilience of obfuscated models to fine-tuning attacks.  In practice, one may try several different simulation networks and choose one with the best performance and the worst fine-tuning ability as the final solution.

\subsection{Discussion}
Besides the cost and fine-tuning ability which we have already evaluated, a competent obfuscation approach should also perform well concerning information leakage, resilience, and scalability.  Below, we discuss the performance of our obfuscation approach in these aspects.

\subsubsection{Information Leakage}
Our obfuscation approach conceals the internal structures of the feature extraction network.  We simulate the entire feature extractor of a CNN model with a simulation network.  In this way, the obfuscated model exposes no information about the original feature extraction network expect the interfaces of input and output, including the raw image size and the number of extracted features.  Besides, the fully-connected layer also leaks the number of classes supported.  However, when attackers want to pirate a deep learning model, such information is not very critical.  The most important information lies in the design of the feature extraction network, which has already been protected by our approach.

\subsubsection{Resilience to Deobfuscation}
We discuss the resilience of our approach in two aspects.  Firstly, can attackers recover the structure of an original network?  The answer is no.  Attackers cannot figure out the original structure of a feature extractor, because such information is not retained within the obfuscated model.  

Secondly, can attackers empower an obfuscated model with respect to learning?  One possible way might be employing another powerful network to simulate the obfuscated model.  However, there are two barriers for attackers to launch such attacks.  One barrier is that attackers should know a powerful network, and another barrier is that they should have high-quality data to train the simulation network.  Note that in our attacker model, we assume the original training data are unavailable to attackers.  Without the training data, attackers can only tune the new kernel parameters with other artificial datasets (\textit{e.g.,} some randomly generated images).

\subsubsection{Scalability}
Currently, we have only evaluated the feasibility of our approach for several CNN models.  Therefore, we claim the effectiveness of our solution for CNN models only.  However, we may extend the idea to other types of neural networks, such as recurrent neural networks.  This is a direction of our future work.  

For CNN models, the performance of our obfuscation approach could be related to the complexity of a target model, such as the depth of the network to simulate and the knowledge it contains.  In this paper, we have verified that our approach can effectively obfuscate a well-trained DenseNet-121 model, which is already very deep with more than a hundred layers.  Therefore, our simulation-based approach is very promising to obfuscate different convolutional neural networks.  Besides, several recent processes (\textit{e.g.,}~\cite{han2015learning,han2016deep}) achieved in model compression area also adopt simulation-based ideas, and their results also coincide the scalability of our approach.  

However, our approach may not be able to obfuscate a model efficiently if it is already very efficient.  For example, we can hardly obfuscate a compressed model with no cost, because such models memorize knowledge efficiently with only a small number of parameters.  In that case, we may apply obfuscation techniques on an uncompressed model first and then compress the obfuscated model.

\section{Related Work}~\label{sec:related_work}
Now we discuss existing papers related to our work.  Our work studies model piracy threats and the deep learning obfuscation problem.  To our best knowledge, it is a pilot study in this area.  There are other investigations which also focus on the security issues of deep learning models (\textit{e.g.,}~\cite{nguyen2015deep,tramer2016stealing,hitaj2017deep}, but their attack models and purposes are very different from ours.

We obfuscate deep learning models via a simulation-based approach.  The similar approaches are also employed in model compression work (\textit{e.g.,}~\cite{lecun1990optimal,hinton2014distilling,denton2014exploiting,han2015learning,han2016deep}).  Some investigations compress a well-trained model without changing its structure, such as network pruning~\cite{lecun1990optimal,han2015learning}, and quantization~\cite{han2016deep}.  Network pruning compresses well-trained CNN models by prune the less important connections and weights.  A quantization-based approach employs fewer bits for each parameter of the model.  They all show that the compressed models after fine-tuning suffer no declination of accuracy.

There also other compression approaches which intend to compress the structure of the network.  They try to distill the knowledge learned by the large and cumbersome network, and install the knowledge into another small model~\cite{hinton2014distilling}.  FitNets~\cite{romerofitnets} is a representative work in this area.  To compress a model, FitNets simulates a deep learning model with another network that is deeper and thinner.  Our work, on the other hand, attempts to simulate a model with a shallow network.  This is because the purpose of our paper is different from model compression.  As a result, we may have different priorities when designing the simulation networks.  Specifically, we purposely hide the structure of a model and degrade its fine-tuning ability.  Therefore, employing a deeper simulation network is not our prior choice, because a deep network is likely to have good learning abilities.  On the other hand, model compression aims to compress a model with high efficiency, while security is not a concern.  FitNets proposes a hint-based training approach to train the simulation networks.  Our joint training approach also incorporates the idea, while it is an enhancement of joint-based training.

In brief, model compression is a technique orthogonal to ours.  Although our approach and existing model compression investigations share some common technical background concerning network simulation, they are different in both purposes and detailed designs.  We may further employ model compression techniques to compress an obfuscated model in our future work.

\section{Conclusion}~\label{sec:conclusion}
In this paper, we have discussed the piracy threats to deep learning models, and we propose to obfuscate such models before releasing them to clients.  To our best knowledge, this paper is a first attempt to investigate the deep learning obfuscation problem.  In particular, we have discussed the concept of deep learning obfuscation and the performance metrics for evaluating a competent obfuscation solution.  Moreover, we propose a structural obfuscation approach for obfuscating CNN models.  Our obfuscation approach simulates the feature extractor of a CNN model with a shallow and sequential convolutional block.  We train the simulation network incorporating a novel joint training method and a recursive training method.  We have verified the effectiveness of our approach with three prevalent CNNs, and the results show that we can obfuscate their structures with no declination of accuracy.  Furthermore, the obfuscated models can be even more efficient than their original models.  As a security benefit, our obfuscated model leaks no critical information about internal structure of the original CNN network.  In the meanwhile, by choosing a shallow simulation network with a poor fine-tuning ability, the obfuscated model can be resilient to parameter piracy attacks.

\bibliography{onn}
\bibliographystyle{abbrv} 

\end{document}